\documentstyle[aps,twocolumn,prl,tighten,flushrt]{revtex} 

\input epsf
\def\plotone#1{\centering \leavevmode
\epsfxsize= 1.0\columnwidth \epsfbox{#1}}

\def\etal{{et al.}}
\def\apjl{Astrophys. J. Lett.}

\def\be{\begin{equation}}
\def\ee{\end{equation}}
\def\bea{\begin{eqnarray}}
\def\eea{\end{eqnarray}}

\def\cmm2{{\,\rm cm^{-2}}}
\def\cm2{{\,{\rm cm}^2}}
\def\cmm3{{\,{\rm cm}^{-3}}}
\def\gcmm3{{\,{\rm g\,cm^{-3}}}}

\def\fun#1#2{\lower3.6pt\vbox{\baselineskip0pt\lineskip.9pt
  \ialign{$\mathsurround=0pt#1\hfil##\hfil$\crcr#2\crcr\sim\crcr}}}

\def\muK{\mu {\rm K}}

\def\etal{{et al.}}

\def\p3m{P$^3$M}

\def\fun#1#2{\lower3.6pt\vbox{\baselineskip0pt\lineskip.9pt
  \ialign{$\mathsurround=0pt#1\hfil##\hfil$\crcr#2\crcr\sim\crcr}}}

\begin{document}
\twocolumn[\hsize\textwidth\columnwidth\hsize\csname @twocolumnfalse\endcsname
\draft
\title{Characterizing the Peak in the \\
Cosmic Microwave Background Angular Power Spectrum}
\author{Lloyd\ Knox$^1$ and Lyman Page$^2$}
\address{$^1$ Department of Astronomy and Astrophysics\\
University of Chicago, 5640 S. Ellis Ave., Chicago, IL 60637, USA}
\address{$^2$ Department of Physics\\
Princeton University, Princeton, NJ, USA}
\date{\today}
\maketitle

\begin{abstract}
  A peak has been unambiguously detected in the cosmic microwave
  background (CMB) angular spectrum.  Here we characterize its
  properties with fits to phenomenological models.  We find that the
  TOCO and BOOM/NA data determine the peak location to be in the range
  175--243 and 151--259 respectively (both ranges 95\% confidence) and
  determine the peak amplitude to be between $\approx70~{\rm and}~90$
  $\muK$.  By combining all the data, we constrain the full-width at
  half-maximum to be between 180 and 250 at 95\% confidence.  Such a
  peak shape is consistent with inflation-inspired flat, cold dark
  matter plus cosmological constant models of structure formation with
  adiabatic, nearly scale-invariant initial conditions. It is
  inconsistent with open and defect models.
\end{abstract}
] 

{\parindent0pt\it Introduction.}  If the adiabatic cold dark matter
(CDM) models with scale-invariant initial conditions describe our
cosmogony, then an analysis of the anisotropy in the CMB can reveal the
cosmological parameters to unprecedented accuracy
\cite{forecast}. A number of studies have aimed at
determining, with various prior assumptions, a subset of the $\sim 10$
free parameters that affect the statistical properties of the CMB
\cite{paramest,dodknox99}.  The parameter most robustly determined
from current data is $\Omega$, the ratio of the mean matter/energy
density to the critical density (that for which the mean spatial
curvature is zero).  These investigations show that
$\Omega$ is close to one. This result, combined with other
cosmological data, implies the existence of some smoothly distributed
energy component with negative pressure such as a cosmological constant.

A weakness of previous approaches \cite{paramest,dodknox99} is that
the conclusions depend on the validity of the assumed model.  In this
{\it Letter} we take a different tack and ask what we know independent
of the details of the cosmological model.  We find the peak location,
amplitude and width are consistent with those expected in adiabatic
CDM models.  Furthermore, as $l_{\rm peak} \simeq 200 \Omega^{-1/2}$
in these models, the observed peak location implies $\Omega \simeq 1$.
The determination of the peak location is robust; it does not depend
on the parametrization of the spectrum, assumptions about the
distribution of the power spectrum measurement errors, nor on the
validity of any one data set.  The model-dependent determinations of
$\Omega$ are further supported by the {\it inconsistency} of the data
with competing models, such as topological defects, open
models with $\Omega < 0.4$, or the simplest isocurvature models.

{\parindent0pt\it The Data.}  
The last year of the 1000's saw new
results from MSAM\cite{wilson99}, PythonV\cite{coble99}, MAT/TOCO
\cite{toco,tocoexplained}, 
Viper\cite{peterson}, CAT\cite{bak99}, IAC\cite{dicker99}
and BOOM/NA\cite{mauskopf}, all of which have bearing on the properties of the 
peak.  These results are plotted in Fig. 1. We have known for several
years that there is a rise toward towards $l=200$ 
but it is now clear that the spectrum also falls significantly towards
$l=400$.

For all the medium angular scale experiments, the largest systematic
effect is the calibration error which is roughly 10\% for each.
Contamination from foreground emission is also important and not yet
fully accounted for in some experiments ({\it e.g.} TOCO). A
correction for this contribution, for which $\delta T_l \sim
l^{-1/2}$, will affect the amplitude of the peak though will not
strongly affect its position. Thorough analyses by the
MSAM\cite{cheng} and PYTHON\cite{coble99} teams show that the level of
contamination in those experiments was $<3\%$.

The three experiments that have taken data that span the peak are
MSAM, TOCO, and BOOM/NA.  All
experiments exhibit a definite increase over the Sachs-Wolfe plateau
though the significance of a feature based on the data alone, e.g. a
peak, differs between experiments.  We may assess the detection of a
feature by examining the deviation from the best fit flat line,
$\overline{\delta T}$.  For the three MSAM points, we find 
$\overline{\delta T}=46\pm 4.9~\mu$K with a reduced $\chi^2$ of 0.43 
(Probability to
exceed, $P_{>\chi^2} = 0.65$. The calibration error is not included.).  
Thus, no feature is detected with these data alone though
there is a clear increase over DMR\cite{DMR}.  
For the seven BOOM/NA points, we
find $\overline{\delta T}=55.3\pm 4.2~\mu$K with a reduced $\chi^2$ of 1.94
($P_{>\chi^2} = 0.05$, assuming the data are anti-correlated at
the 0.1 level\cite{mauskopf}). 
For the ten TOCO
points, $\overline{\delta T}=69.3\pm 2.7~\mu$K with a reduced $\chi^2$ of 4.86
($P_{>\chi^2} <10^{-5}$)
Calibration errors will not
change $\chi^2/\nu$, however a correction for foreground emission will 
have a slight effect.
Though we examine all data in the following, we focus particularly on
BOOM/NA and TOCO because of their detections of a feature.

{\parindent0pt\it Fits to Phenomenological Models.} 
To characterize the peak amplitude and location we fit the
parameters of two different phenomenological models.  
For the first, we start with the best fit DK99\cite{dodknox99} adiabatic
CDM model, $\delta T_l^{DK}$, and form
$\delta T_l = (\delta T_l^{DK}-\delta T_{l=10}^{DK})\alpha + \delta T_{l=10}^{DK}$
by varying $\alpha$, and then stretching in $l$\cite{barth96}. We characterize
each stretching with the peak position and peak amplitude. 
This method has the virtue that the resulting
spectra resemble adiabatic models and so if one assumes that
these models describe Nature, then these results are the
ones to which we should pay the most attention.

\begin{figure}[htbp]
  \plotone{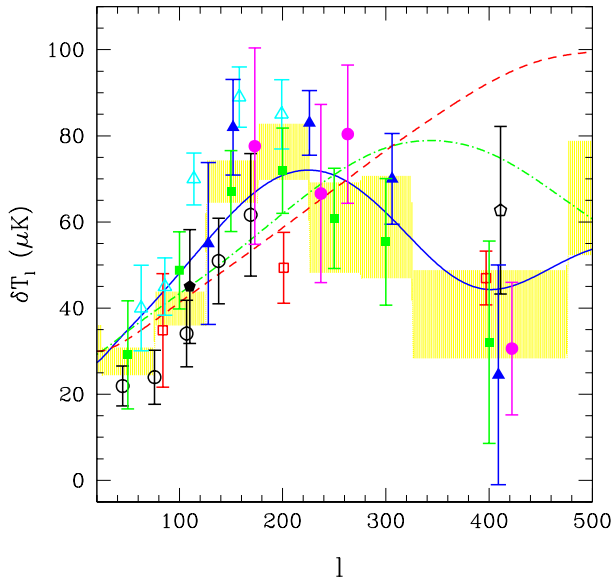} 
    \caption{Bandpowers from TOCO97 (cyan open triangles), 
TOCO98 (blue filled triangles), BOOM/NA (green filled squares),
MSAM (red open squares), CAT (black open pentagon), IAC (black 
filled pentagon), PyV (black open circles) and Viper
(green filled circles). The y-axis is 
$\delta T_l \equiv \sqrt{l(l+1)C_l/(2\pi)}$ where $C_l$ is the
angular spectrum. The models
are, peaking at left to right, the best fit models
of \protect\cite{dodknox99} for $\Omega=1$, $\Omega=0.4$ and $\Omega=0.2$.  
The $\Omega=1$ model has a mean density of non-relativistic matter, 
$\Omega_m=0.31$, a cosmological constant density of $\Omega_\Lambda=0.69$,
a baryon density of $\Omega_b=0.019h^{-2}$ \protect\cite{BNTT99}, a Hubble
constant of $H_0 =65 \ {\rm km/sec/Mpc}$, 
an optical depth to reionization
of $\tau = 0.17$ and a power spectrum power-law index of $n=1.12$, where
$n=1$ is scale-invariant.
The shaded areas are the results of 
fitting the power in 14 bands of $l$ to all the data (from 1999 and previous years) as in \protect\cite{bjk98}.
Many of the bands are at low  $l$ and cannot be discerned on this plot.
Calibration errors are not shown though are included in the best fit.
}
\label{fig:data}
\end{figure}

Our second model for $\delta T_l^2$ is a Gaussian:
$\delta T_l^2 = A^2 \exp\left(-\left(l-l_c\right)^2/(2\sigma_l^2)\right)$.
Depending on the width, this spectrum can look
very much like, or unlike, the spectra of adiabatic models  
\cite{gaussmod}.
We view this versatility as a virtue since we are interested 
in a characterization of the peak which is independent of physical models.

We fit to these phenomenological models in two ways. For the stretch
model, we examine the $\chi^2$ of the residuals between the 
published data and each model.
The widths of the window functions are ignored and we assume the data
are normally distributed in $\delta T_l$ with a dispersion
given by the average of the published error bars (GT in Table 1). This is an admittedly
crude method but it works well because the likelihoods as a function of
$\delta T_l$ are moderately well approximated by a normal distribution.

For both the Gaussian shape and the stretch model, we also perform
the full fit as outlined in BJK\cite{bjk98} (RAD in Table 1). 
For the Gaussian shape model, the
constraints on the amplitude and
location are given below after marginalization over the width $\sigma_l$.
In all fitting, we ignore the experiments that are affected by $l< 30$
(DMR, FIRS\cite{firs} and Tenerife\cite{tenerife}) because
we want the parameters of
our Gaussian to be determined by behavior in the peak region.

\vskip 0.15in
\begin{tabular}{cccccccc}
\hline
\hline
Data & Model & Fit & $N/\nu$ & $\chi^2/\nu$ & $P_{>\chi^2}$ &$l_{peak}$ &$\delta T_{peak}$ \\
&  &  &  &   &   &  & $~\mu$K \\
\hline
All  & G & Rad           & 58/55 & 1.25 &0.10 &$229 \pm 8.5$ & 78 \\ 
T & G & Rad              & 10/7 & 0.41 &0.89 & $206 \pm 16$ & 95 \\
T & S & GT               & 10/8 & 0.94 & 0.48 & $214\pm 14$ & 88 \\
T & S & Rad              & 10/8 & 0.84 & 0.57 & $209\pm 17$ & 92 \\
B & G & Rad              & 7/4  & 0.19 &0.94&$208 \pm 21$ & 69   \\
B & S & GT               & 7/5  & 0.39 & 0.85 & $215\pm 24$ & 69 \\
B & S & Rad$_0$      & 7/5  & 0.23 & 0.95 & $205\pm 27$ & 72 \\
B & S & Rad$_\infty$ & 7/5  & 0.39 & 0.85 & $206\pm 26$ & 68 \\
P & G & Rad          & 33/30  & 1.13 & 0.28 & $262\pm 24$ & 68 \\
\hline
\end{tabular}
\footnotemark{ALL stands for all publically available data sets (except
for VIPER which was not used because of unspecified 
point-to-point correlations), the T
is for the TOCO data, the B for BOOM/NA and the P is for ``Previous'',
meaning all data prior to BOOM/NA and TOCO.}
\footnotemark{G and S are for the Gaussian shape and stretch methods
respectively}
\footnotemark{$N$ is the number of data points and $\nu$ the degrees of freedom.}
\footnotemark{Rad$_0$ and Rad$_\infty$ corresponds to log
normal and normal distributions for the likelihood respectively.}
\vskip 0.15in

The main thing to notice in the Table is that the position of the peak
is robustly determined by {\it either} TOCO or BOOM/NA to be in the
range 185 to 235, regardless of the method. For the quoted errors, we
have marginalized over all parameters except the position.  The
peak amplitudes are subject to change as
there is some dependence on the model parametrization and the
foreground contamination has not been thoroughly assessed.

We account for the calibration uncertainty through a convolution
of the likelihood of the fits with a normal distribution of the fractional
error \cite{ganga97,bjk98}. BOOM/NA, TOCO97 and TOCO98 have
calibration uncertainties of 
8\%, 10.5\% and 8\% respectively.  However, 5\% of this is
due to uncertainty in the temperature of Jupiter and therefore,
assuming that these uncertainties add in quadrature, we get
$\sigma_{\rm Jup} = 0.05$, $\sigma_{T97}=0.092$, $\sigma_{T98}=0.062$ 
and $\sigma_{B97} = 0.062$.  We then find, for TOCO, that the full likelihood
in $\delta T_l$ and $l$ is given by

\bea
\label{eqn:margi}
L(l_c,\delta T_l) &=& \int d\sigma_l du_{\rm Jup} du_{T97} du_{T98} 
L_{T97}(l_c,\delta T_lu_{\rm Jup}u_{T97},\sigma_l)  \nonumber \\
& & \times L_{T98}(l_c,\delta T_lu_{\rm
Jup}u_{T98},\sigma_l)P_G(u_{97}-1;\sigma_{T97})
  \nonumber \\
& & \times P_G(u_{T98}-1;\sigma_{T98})
P_G(u_{\rm Jup}-1;\sigma_{\rm Jup}) 
\eea
where $P_G(x;\sigma)= \exp\left(-x^2/(2\sigma^2)\right)/ \sqrt{2\pi\sigma^2}$,
$u$ is integrated from 0 to $\infty$
and, e.g., $L_{T97}(l_c,\delta T_l,\sigma_l) = \exp(-\chi^2/2)$
where $\chi^2$ is evaluated on a grid of $\delta T_l^2$, $l_c$ \&
$\sigma_l$ using RADPACK\cite{radpack} as discussed in BJK.
We get similar results for TOCO when simply using a combined
total calibration error of 8.5\%.

For the Gaussian model we can also marginalize over $A$ and $l_c$ to
place 95\% confidence bounds on the width:
$75 < \sigma_l < 105$ for ALL, $50 < \sigma_l < 105$ for TOCO
and $55 < \sigma_l < 145$ for BOOM/NA.

\begin{figure}[bthp]
  \plotone{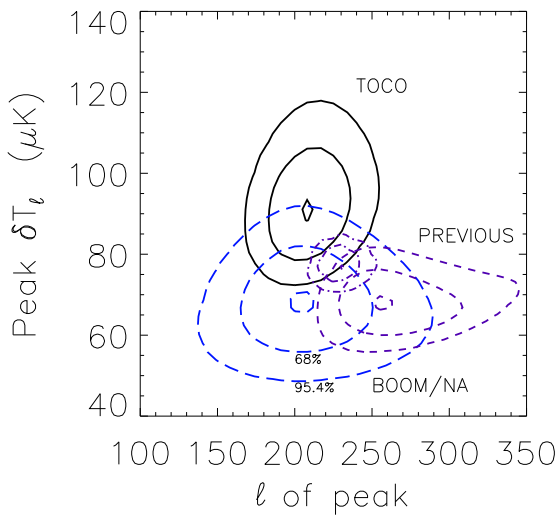}
  \caption{Likelihood contours for $l$ vs $\delta T_l$ for the position
of the peak. For BOOM/NA and TOCO, we use the stretch 
method using RADPACK\protect\cite{radpack} and include
the calibration error. 
For Previous and ALL (tightest contours) we have {\it not} used generalizations
of Eq.~\ref{eqn:margi}, but instead have fixed the calibration 
parameters. All contour levels correspond to 5\%, 68\%,
and 95\% enclosed, or roughly the peak, $1\sigma$,
$2\sigma$. }
\label{fig:Avl}
\end{figure}

Are the
data in Fig 1 consistent? DK99 found that the best-fit model, given all the
data at the time, had a $\chi^2$ of 79 for 63 degrees of freedom,
which is exceeded 8\% of the time.  Here we see that the $\chi^2$ for
the fit of the Gaussian model is 69 for 55 degrees of freedom, which
is exceeded 10\% of the time.  We conclude that, although there may
well be systematic error in some of these data sets, we have no
compelling evidence of it.  However, we take caution from the fact
that we had to adjust the calibration parameters from their nominal
values to their best-fit values in order to reduce the $\chi^2$ to 69.
Left at their nominal values with calibration uncertainty ignored, the
data are not consistent with each other.  Thus we believe that the
compilation results are perhaps less reliable than those for either
BOOM/NA or TOCO.

{\parindent0pt\it Implications for Physical Models.} 
Flat, adiabatic, nearly scale-invariant models 
have similar peak properties to those of our best-fit 
phenomenological models.  Most importantly the peak location,
as determined by three independent data sets (``Previous'',
TOCO, BOOM/NA), is near $l \simeq 210$, as expected.  Depending on
the data set chosen, the amplitude is higher than expected but
can easily be accommodated, within the uncertainties, 
with a cosmological constant.
Combining all the data, there is a preference 
for $l_{\rm peak} > 210$ which suggests a cosmological constant\cite{pascos}
(at $h=0.65$, $l_{\rm peak}$ goes
from 200 at $\Omega_\Lambda=0$ to 220 at $\Omega_\Lambda=0.7$).
However, this result is not seen in any individual data set.

A good approximation to the first peak in the DK99 best-fit model is
given by the Gaussian model with $\sigma_l=95$.  From the $\sigma_l$
constraints quoted earlier we see that the data have no significant
preference for peaks that are either narrower or broader than those in
inflation-inspired CDM models.

A general perturbation is a combination of adiabatic and isocurvature
perturbations.  Adiabatic perturbations are such that at each point in
space, the fractional fluctuations in the number density of each
particle species is the same for all species.  Isocurvature
perturbations are initially arranged so that, despite fluctuations in
individual species, the total energy density fluctuation is zero.
Given multiple components, there
are a number of different ways of maintaining the isocurvature condition.
Below we assume the isocurvature condition
is maintained by the dark matter compensating everything else.

Isocurvature initial conditions result in shifts to the CMB power
spectrum peak locations.  For
a given wavenumber, the temporal phase of oscillations in the
baryon-photon fluid depends on the initial relation between
the dark matter and the fluid.  
Those waves with oscillation frequencies
such that they hit an extremum at the time of last-scattering in
the adiabatic case, will hit a null in the isocurvature case\cite{husugwhite}.
The effect on the first peak is a shift from 
$l \simeq 200\Omega^{-1/2}$ to $l \simeq 350\Omega^{-1/2}$.
Given the observation of  $l_{\rm peak} \simeq 210$, simple isocurvature
models require $\Omega > 2$---which is inconsistent with
a number of observations\cite{montypython}.

Critical to the Doppler peak structure, in either adiabatic or
isocurvature models, is the temporal phase coherence for 
Fourier modes of a given wavenumber\cite{coherence}.  In topological defect
models, the continual generation of new perturbations by
the non-linear evolution of the defect network destroys this
temporal phase coherence and the acoustic peaks blend into a
broad hump which is wider and peaks at higher $l$ than
the observed feature.  

One can make defect model power spectra with less power at $l=400$
than at $l=200$ with ad-hoc modifications to the standard 
ionization history\cite{WBA}.  But even for these models the drop
is probably not fast enough\cite{Albrechtpascos}.  
The contrast between the power at $l=200$ and
$l=400$ is a great challenge for these models.

There are scenarios with initially isocurvature conditions
that can produce CMB power spectra that look much
like those in the adiabatic case.  This can be done by adding
to the adiabatic fluctuations (of photons, neutrinos, baryons and
cold dark mater) another component, with a non-trivial stress
history, which maintains
the isocurvature condition\cite{postmodernisocurv}.  

{\parindent0pt\it Conclusions.} Our phenomenological 
models have allowed for rapid, model-independent,
investigation of the consistency of CMB datasets, and of the
robustness of the properties of the peak in the CMB power spectrum.
The peak has been observed by two different instruments, and can be
inferred from an independent compilation of other data sets.  The
properties of this peak are consistent with those of the first peak in
the inflation-inspired adiabatic CDM models, and inconsistent with a
number of competing models, with the possible exception of the more
complicated isocurvature models mentioned above.  It is perhaps
instructive that where the confrontation between theory and
observation can be done with a minimum of theoretical uncertainty, the
adiabatic CDM models have been highly successful.

\acknowledgements 
LK wishes to thanks S. Meyer and M. Tegmark for useful conversations and 
is supported by the DOE, NASA grant NAG5-7986 and NSF grant OPP-8920223.
LP wishes to thank MAT/TOCO team members Mark Devlin, Randy Dorwart,
Rob Caldwell, Tom Herbig, Amber Miller, Michael Nolta, Jason Puchalla, 
Eric Torbet, \& Huan Tran,
for insights and encouragement, and Chuck Bennett for comments on
an earlier version of this work. LP is supported by NSF grant PHY
96-00015 and NASA grant NAS5-96021.


\begin{thebibliography}{ucsc}
\bibitem{tocoexplained} The Mobile Anisotropy Telescope
(MAT), while situated on Cerro Toco, Chile, produced two data sets using
the same instrument and observing
strategy, TOCO97 \& TOCO98. We refer to these collectively as TOCO.
\end{thebibliography}
\end{document}